\documentclass{article}
\usepackage{ijcai21}

\usepackage{times}
\usepackage{soul}
\usepackage{url}
\usepackage{pgfplots}
\usepackage[hidelinks]{hyperref}
\usepackage[utf8]{inputenc}
\usepackage[small]{caption}
\usepackage{graphicx}
\usepackage{amsmath}
\usepackage{amsthm}
\usepackage{booktabs}
\usepackage{algorithm}
\usepackage{algorithmic}
\usepackage{subfigure}

\title{MG-DVD: A Real-time Framework for Malware Variant Detection Based on Dynamic Heterogeneous Graph Learning}

\author{
Chen Liu$^{1,2}$\and
Bo Li$^{1,2}$\footnote{Corresponding Author}\and
Jun Zhao$^{1,2}$\and
Ming Su$^{1,2}$\And
Xu-Dong Liu$^{1,2}$\footnote{Corresponding Author}\\
\affiliations
$^1$School of Computer Science and Engineering, Beihang University, Beijing, China\\
$^2$Beijing Advanced Innovation Center for Big Data and Brain Computing, Beihang University, China\\
\emails
\{liuchen, libo, zhaojun, suming, liuxd\}@act.buaa.edu.cn
}
\pgfplotsset{compat=1.14}

\begin{document}

\maketitle

\begin{abstract}
Detecting the newly emerging malware variants in real time is crucial for mitigating cyber risks and proactively blocking intrusions. In this paper, we propose MG-DVD, a novel detection framework based on dynamic heterogeneous graph learning, to detect malware variants in real time. Particularly, MG-DVD first models the fine-grained execution event streams of malware variants into dynamic heterogeneous graphs and investigates real-world meta-graphs between malware objects, which can effectively characterize more discriminative malicious evolutionary patterns between malware and their variants.
Then, MG-DVD presents two dynamic walk-based heterogeneous graph learning methods to learn more comprehensive representations of malware variants, which significantly reduces the cost of the entire graph retraining.
As a result, MG-DVD is equipped with the ability to detect malware variants in real time, and it presents better interpretability by introducing meaningful meta-graphs.
Comprehensive experiments on large-scale samples prove that our proposed MG-DVD outperforms state-of-the-art methods in detecting 
malware variants in terms of effectiveness and efficiency. 
\end{abstract}

\section{Introduction}
Malware has become one of the largest catastrophes endangering information systems, which consistently permeates and attacks information systems to steal sensitive information, take control of the target system, and collect ransom~\cite{gandotra2014malware}. Purifying the network environment and keeping information systems away from malware attacks has become a serious challenge faced by security communities and researchers~\cite{ye2019out}.
\begin{figure}[h]  
    \centering 
    \includegraphics[width=7.1cm]{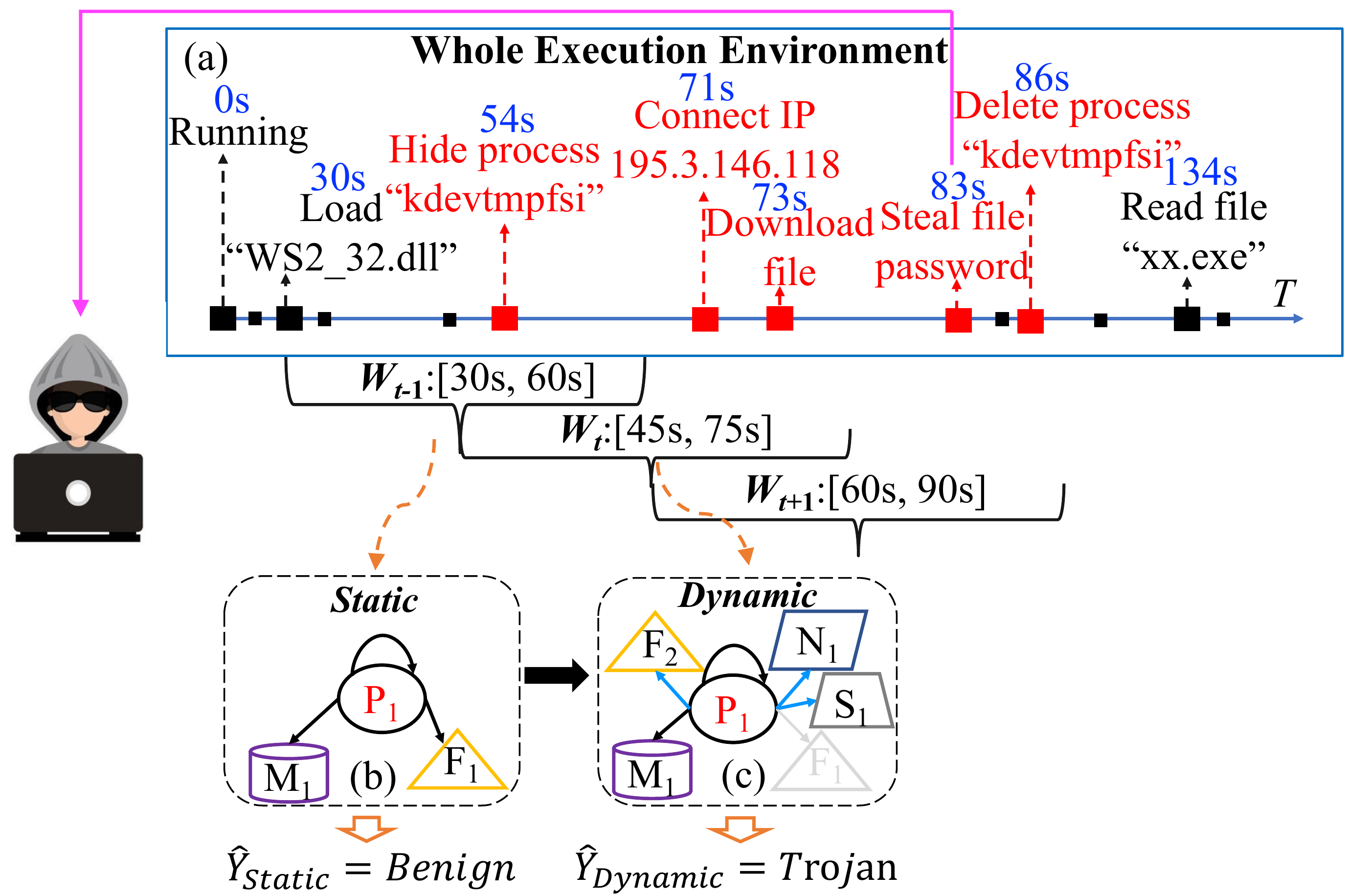}  
    \caption{Illustration of dynamic malware variant detection: (a) depicts execution events stream of ``\emph{ kdevtmpfsi}", in which red rectangles and black rectangles represent malicious behaviors and normal behaviors, respectively; (b) represents the static-based detection method~(e.g., MatchGNet~\protect\cite{DBLP:conf/ijcai/WangCYLNTGLCY19}), which only models the execution events at a specific time~(e.g., \emph{W$_{t-1}$}) into a static heterogeneous graph but cannot investigate newly added events in real time, result in misjudging malicious behaviors as benign.
    (c) is our proposed MG-DVD, which dynamically updates from (b) by adding the blue directed edges~(e.g., \emph{P$_{1}\rightarrow F_{2}$}, \emph{P$_{1}\rightarrow N_{1}$}, and \emph{P$_{1}\rightarrow S_{1}$}) and removing the grey directed edges~(e.g., \emph{P$_{1}\rightarrow F_{1}$}). MG-DVD can dynamically learn more comprehensive representations of malware and capture the evolutionary patterns between malware and variants to detect malicious behaviors in real time~(before 75s).}
    \label{fig:example}
\end{figure}

During the last decade, a large volume of malware detection approaches has been proposed. The existing malware detection methods can be roughly divided into two types: signature-based~\cite{kang2016n,gaviria2017using,raff2018malware,zhang2019feature,singh2020survey} and behavior-based~\cite{pascanu2015malware,sun2016monet,bartos2016optimized,zhang2018sensitive,DBLP:conf/ijcai/WangCYLNTGLCY19}. Particularly, signature-based methods rely on feature engineering to manually extract malware fingerprints from known samples, which hardly identify new malware variants~\cite{ye2019out}. Behavior-based detection methods focus on investigating the independent API sequence rather than considering their interactive call relationships, inevitably resulting in high false positives and expensive time cost~\cite{zhang2020dynamic}.
\begin{figure*}[t]  
    \centering  
    \subfigure[Network schema]{
        \begin{minipage}[t]{0.4\linewidth}\hspace{-10mm}
            \centering  
            \includegraphics[width=0.8\linewidth]{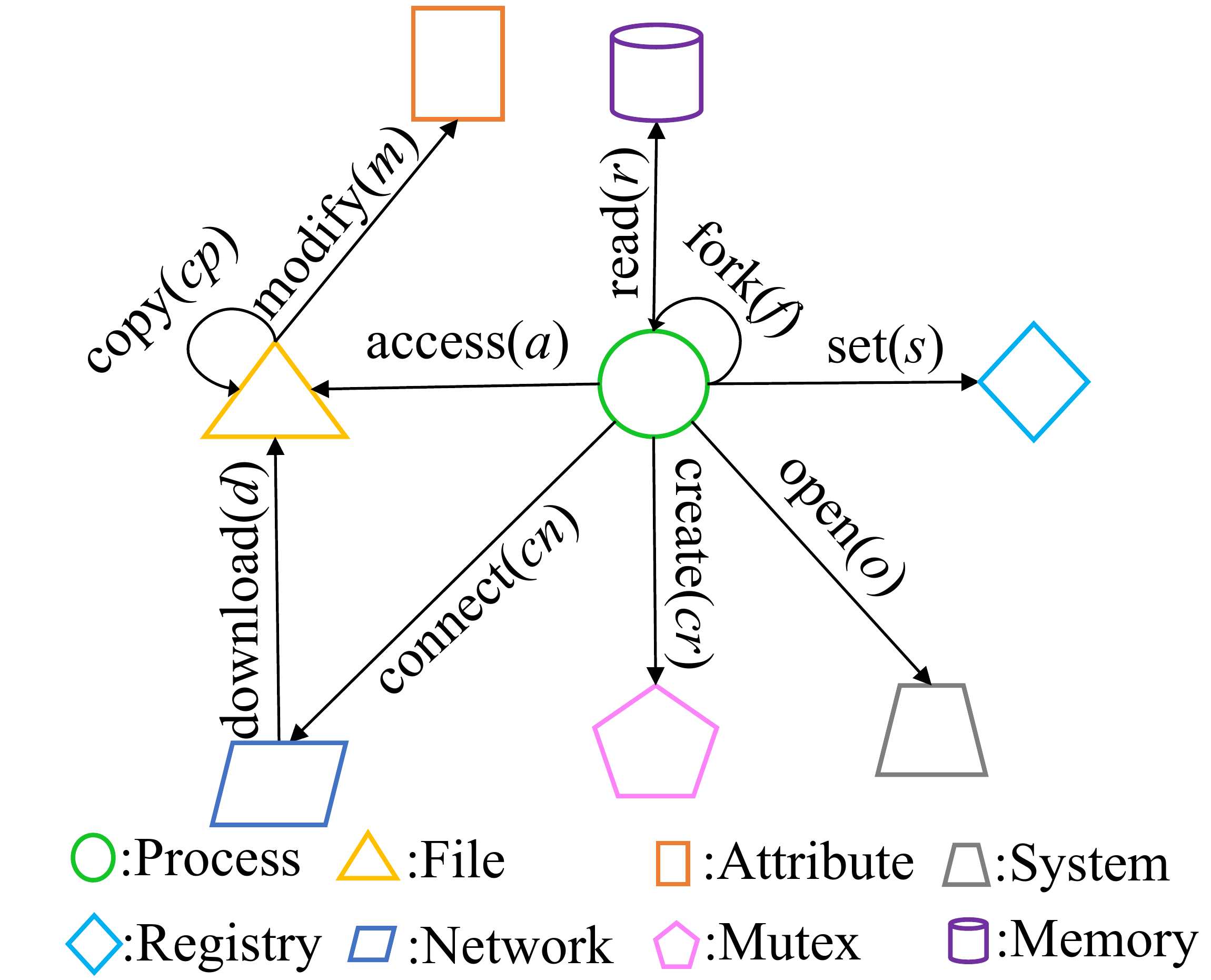}   
        \end{minipage}
}\subfigure[Meta-graphs]{
        \begin{minipage}[t]{0.5\linewidth}\hspace{-10mm}
            \centering     
         \includegraphics[width=1\linewidth]{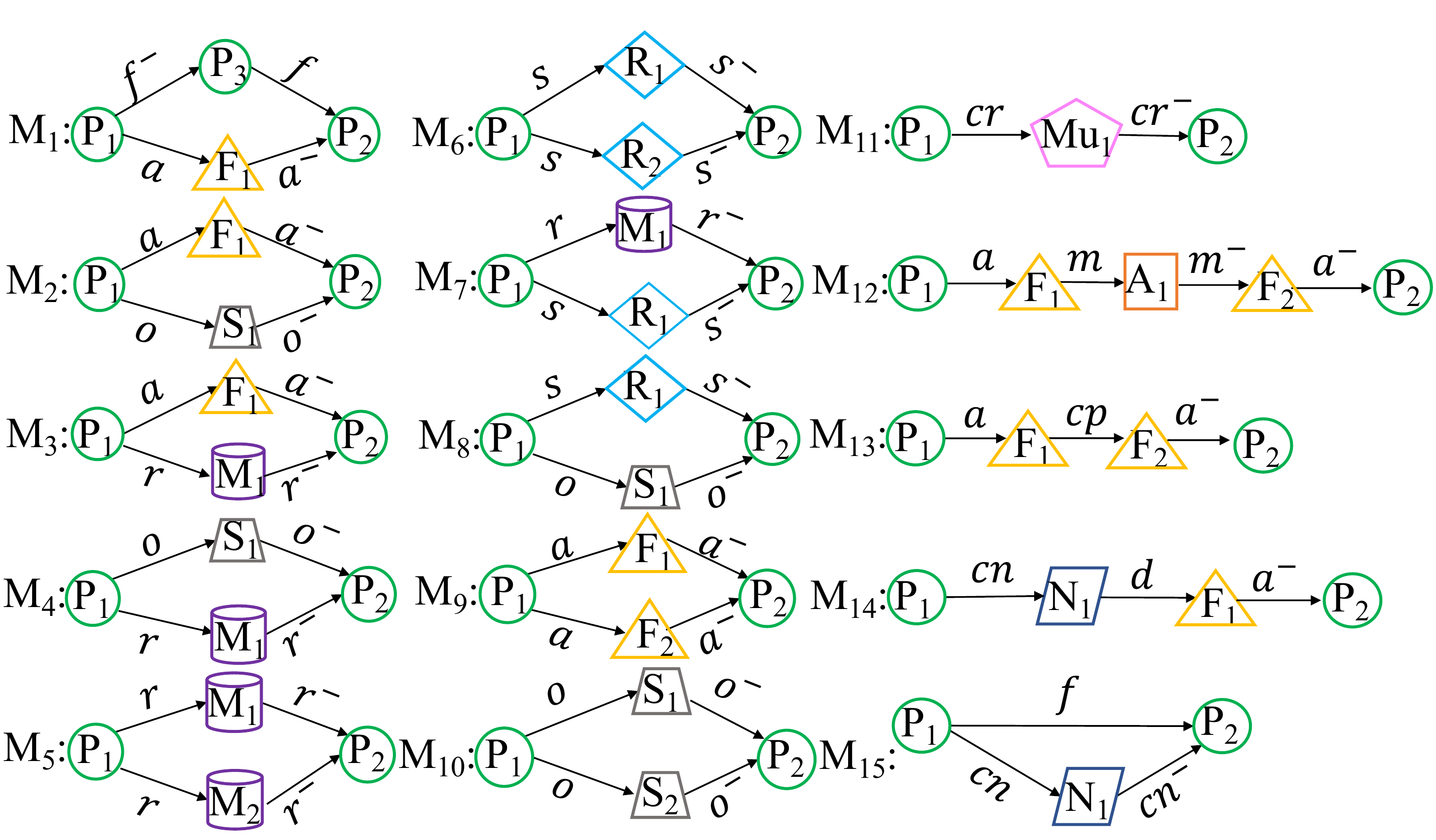}   
        \end{minipage}
    }
    \caption{Network schema and meta-graphs for MG-DVD.} 
    \label{fig:schema} 
\end{figure*}

Indeed, most malware will perform variants over time to evade detection, and their malicious behaviors are triggered in a very short time~\cite{DBLP:conf/ijcai/WangCYLNTGLCY19}. As shown in Figure~\ref{fig:example}, the malicious behaviors of ``{\em kdevtmpfsi}" are submitted between 54s and 86s. However, the existing detection approaches cannot meet the
dynamic detection requirements and misjudge malicious behaviors as benign.
In summary, the existing malware detection approaches expose three major challenges. First, they~\cite{DBLP:conf/ijcai/WangCYLNTGLCY19,chen2018fastgcn,wu2019comprehensive} are inefficient to indiscriminately retrain the entire graph whenever a new node or event is injected, resulting in their incapacity of detecting malware variants in real time. Second, static random walk strategy~\cite{DBLP:conf/ijcai/WangCYLNTGLCY19,fu2017hin2vec,fan2018gotcha} often generates a large search space since long-term walks for all nodes, causing an expensive and unacceptable time cost. Third, feature-based~\cite{kang2016n,gaviria2017using,raff2018malware,zhang2019feature,pascanu2015malware,sun2016monet,bartos2016optimized,zhang2018sensitive} and metapath-based~\cite{DBLP:conf/ijcai/WangCYLNTGLCY19} detection approaches are incapable of characterizing high-order and expressive representations of malware variants.

To tackle these challenges, we present a novel MetaGraph-guided Dynamic Variant Detection~(MG-DVD) framework based on dynamic heterogeneous graph learning, which models execution events stream of the target variant into a series of dynamic heterogeneous graphs and identifies the variant in a real-time and interpretable way. The contributions of this paper are summarized as follows:

\begin{itemize}
  \item
We propose a dynamic detection framework, namely MG-DVD, consisting of two encoders involving DWIUE and CHGAE. MG-DVD utilizes the overlapping information of adjacent sliding windows to dynamically generate graph embedding, 
which provides malware warnings in real time to keep information systems from malicious intrusions.
 \item
MG-DVD implements an efficient dynamic walk strategy, which concentrates on the newly injected nodes in the heterogeneous graph at the target sliding window and learns their node representations by exploring the emerging meta-graphs between them. As a result, MG-DVD is able to effectively alleviate the pressure of search space and time overhead.
\item
MG-DVD presents remarkable interpretability by introducing the meaningful meta-graphs depicting interactive relationships between malware and corresponding variants, which can effectively learn the evolutionary patterns of variants from different malware families.
\item
We conduct extensive experiments on large-scale real-world samples. Experimental results verify that our proposed MG-DVD outperforms the state-of-the-art approaches in detecting new malware variants.
\end{itemize}

\section{Preliminary}
In this section, we recap important definitions used in our work, such as DHGS, network schema, and meta-graph. 

\newtheorem{definition}{Definition}
\begin{definition} \label{definition1}
A {\bf \em dynamic heterogeneous graphs sequence (DHGS)} over the execution events stream of malware variant is a graph set {\bf \em G}=$\{$G$_{1}$, G$_{2}$,\dots, G$_{T}\}$, each G$_{t}$ = (V$_{t}$, E$_{t}$) with an entity type mapping $\phi$: V$_{t}$ $\rightarrow$ A and a relation type mapping $\psi$: E$_{t}$ $\rightarrow$ R, where V$_{t}$
and E$_{t}$ denote the entity set and the relation set of G$_{t}$, respectively, and A and R denote the entity type and relation type, respectively. Among them, $\vert$A$\vert$ $\textgreater$ 1, $\vert$R$\vert$ $\textgreater$ 1. The {\bf \em network schema}~\cite{sun2012mining} of DHGS, denoted as T$_{G}$ = (A, R), is a graph with nodes as entity type from A and edges as relation type from R.
\end{definition}

To characterize high-order semantic information and capture the more discriminative malicious patterns of various malware types, we investigate the real-world meta-graphs for dynamic malware variants detection, defined as below:
\begin{definition} \label{definition3}
A {\bf \em meta-graph}~\cite{zhao2017meta} M is a directed acyclic graph with a single source node n$_{s}$~(i.e., with in-degree 0) and a single target node n$_{t}$~(i.e., with out-degree 0), defined on a DHGS {\bf \em G} with schema T$_{G}$ = (A, R), then a meta-graph can be defined as M=(V$_{M}$, E$_{M}$, A$_{M}$, R$_{M}$, n$_{s}$, n$_{t}$), where V$_{M}\in$~V, E$_{M}\in$~E are constrained by A$_{M}\in$~A and R$_{M}\in$~R, respectively.
\end{definition}
\begin{figure*}[t]  
    \centering 
    \includegraphics[width=14cm]{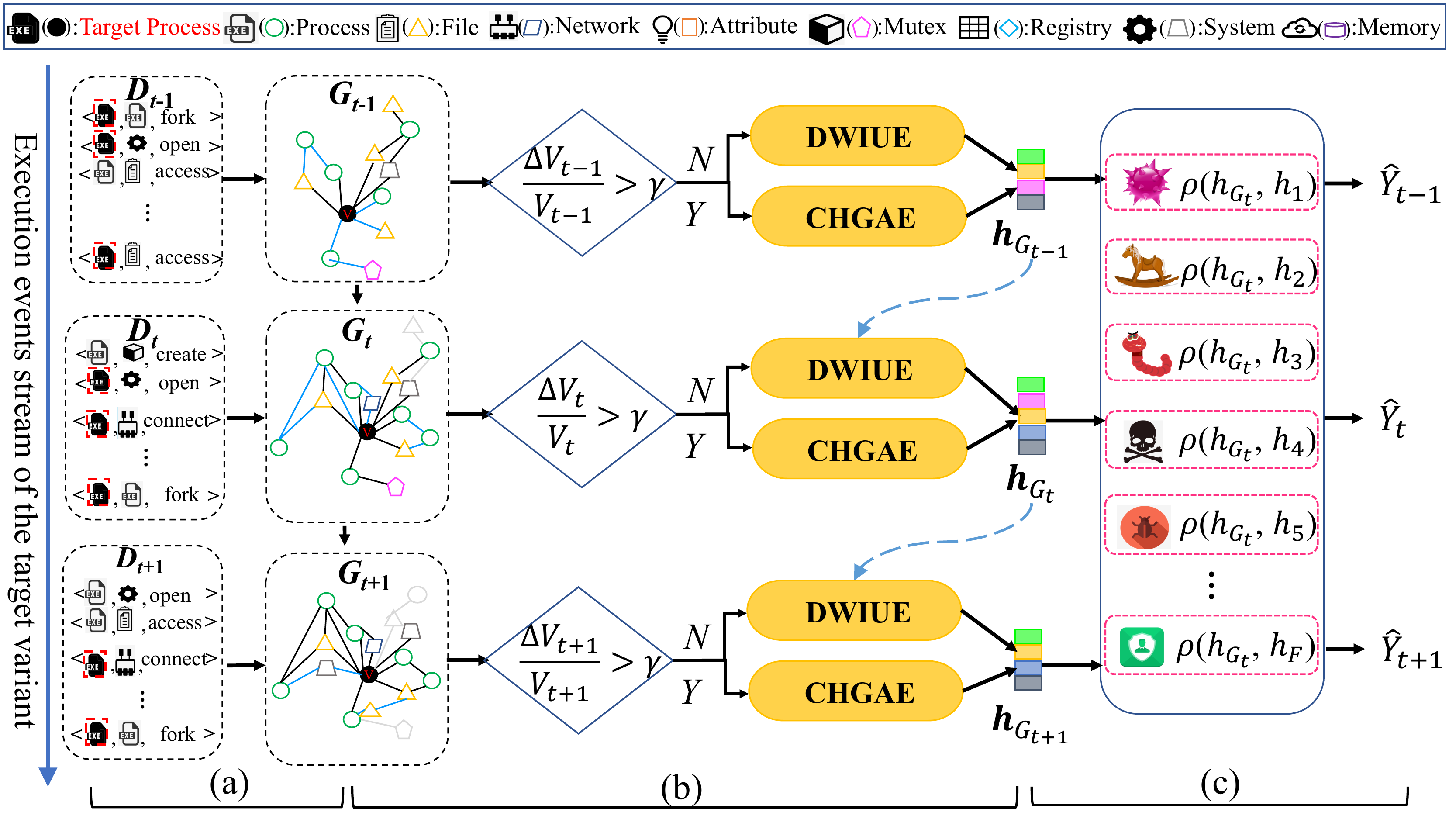}  
    \caption{Framework of MG-DVD: (a) Dynamic heterogeneous graphs constructing: {\em G$_{t}$} is updated based on {\em G$_{t-1}$} and {\em D$_{t}$}, in which the grey nodes and grey solid lines indicate the expired events, and blue solid lines indicate newly-created events (Section 3.1); (b) Dynamic graph learning: generate the graph embedding {\bf \em h$_{G_{t}}$} under each sliding window by encoder DWIUE or CHGAE based on the ratio $\Delta${\em V$_{t}$}/ {\em V$_{t}$} (Section 3.2); 
    (c) Real-time detection: calculate the Pearson correlation coefficients between the graph embedding of the target variant and the graph embeddings of all samples to obtain real-time prediction result $\widehat{Y}_{t}$ (Section 3.3).}
    \label{fig:framework}
\end{figure*}

Particularly, Figure~\ref{fig:schema}(a) demonstrates the schema of MG-DVD, which consists of 8 types of entities and 10 types of relations. With the schema, we first enumerate all meta-graphs of MG-DVD, which start and end with processes~(i.e., {\em P$_{1}$} and {\em P$_{2}$}). Then the frequency of each meta-graph is measured on large-scale samples, and the final meta-graphs are obtained, as shown in Figure~\ref{fig:schema}(b), which hold high-frequency and rich semantics depicting the evolutionary patterns between malware and their variants. 

\section{Framework}
Figure~\ref{fig:framework} shows the framework of MG-DVD, which consists of three major components, detailed below.

\subsection{Dynamic Heterogeneous Graphs Constructing}
MG-DVD models the execution events stream of the malware variant into a {\em DHGS}=$\{${\em G}$_{1}$, {\em G}$_{2}$, \dots, {\em G$_{T}$}$\}$ with different sliding windows~\cite{wang2019three}. As shown in Figure~\ref{fig:framework}, {\em G$_{t}$} can be updated from {\em G$_{t-1}$} as the new execution events are joined, and expired events are removed, and then the heterogeneous graph {\em G$_{t}$} at {\em W}$_{t}$ is obtained, represented as adjacency matrix {\em A$_{t}$}. MG-DVD utilizes overlapping information %involving nodes and links 
at adjacent sliding windows~(e.g., {\em G$_{t-1}$} and {\em G$_{t}$}) to receive richer variant evolution behaviors rather than reconstructing the entire graph under each sliding window, which offers better efficiency. 
Experimentally, the detection performance depends on the sliding window size~(i.e., {\em W}) and sliding step~(i.e., {\em p}), and it is proved that MG-DVD could reach the desired performance when {\em W} is set to 60s and {\em p}=1/2{\em W}. 

\subsection{Dynamic Graph Learning}
Unlike the static frameworks~\cite{chen2018fastgcn,hamilton2017inductive,wu2019comprehensive},
MG-DVD aims to implement a dynamic malware variants detection based on the overlapping information of heterogeneous graphs under adjacent sliding windows. To this end, MG-DVD proposes Dynamic Walk Incremental Updating Encoder~(DWIUE) and Combined Hierarchical Graph Attention Encoder~(CHGAE), which can discriminatively update or retrain the graph embedding at each target sliding window based on the threshold {\em $\Delta V_{t}$}/{\em V$_{t}$}. Particularly, given {\em G$_{t}$} and {\em $\Delta V_{t}$} at sliding window {\em W}$_{t}$, the graph embedding of {\em G$_{t}$} can be generated under the two paradigms. On the one hand, if {\em $\Delta V_{t}$}/{\em V$_{t}$}$\leq\gamma$, MG-DVD utilizes DWIUE to straightly assemble the representation matrix based on the strongly correlated {\bf \em h$_{G_{t-1}}$} and dynamic neighbors; on the other hand, MG-DVD presents CHGAE to partially learn the graph embedding %of the graph at \emph{W}$_{t}$
by discriminatively aggregating newly joined neighbors.
Our presented dynamic walk
strategy only traverses the changed nodes rather than all nodes in the target graph, which markedly alleviates the search space and time cost compared with the static random walk~\cite{fu2017hin2vec,fan2018gotcha}. Definitely, the set of changed nodes on the target graph is denoted as a dynamic node set $\Delta V_{t}$, which is formalized as below:

\begin{equation}
\begin{split}
      \Delta V_{t}=\bigcup\{v_{x}\in V_{t}|(\exists e_{x,y}=(v_{x},v_{y})\notin E_{t-1}) \\
         \vee(\exists e_{x,y}=(v_{x},v_{y})\in E_{t-1}\backslash E_{t})\},\label{eq:no1}
\end{split}
\end{equation}
where {\em v$_{x}$} and {\em v$_{y}$} are nodes in {\em V$_{t}$}, {\em e$_{x,y}$} is the relation between them. {\em E$_{t-1}\backslash$ E$_{t}$} denotes the difference set of {\em E$_{t-1}$} and {\em E$_{t}$}. 
\begin{figure*}[t]  
    \centering 
    \includegraphics[width=14.5cm]{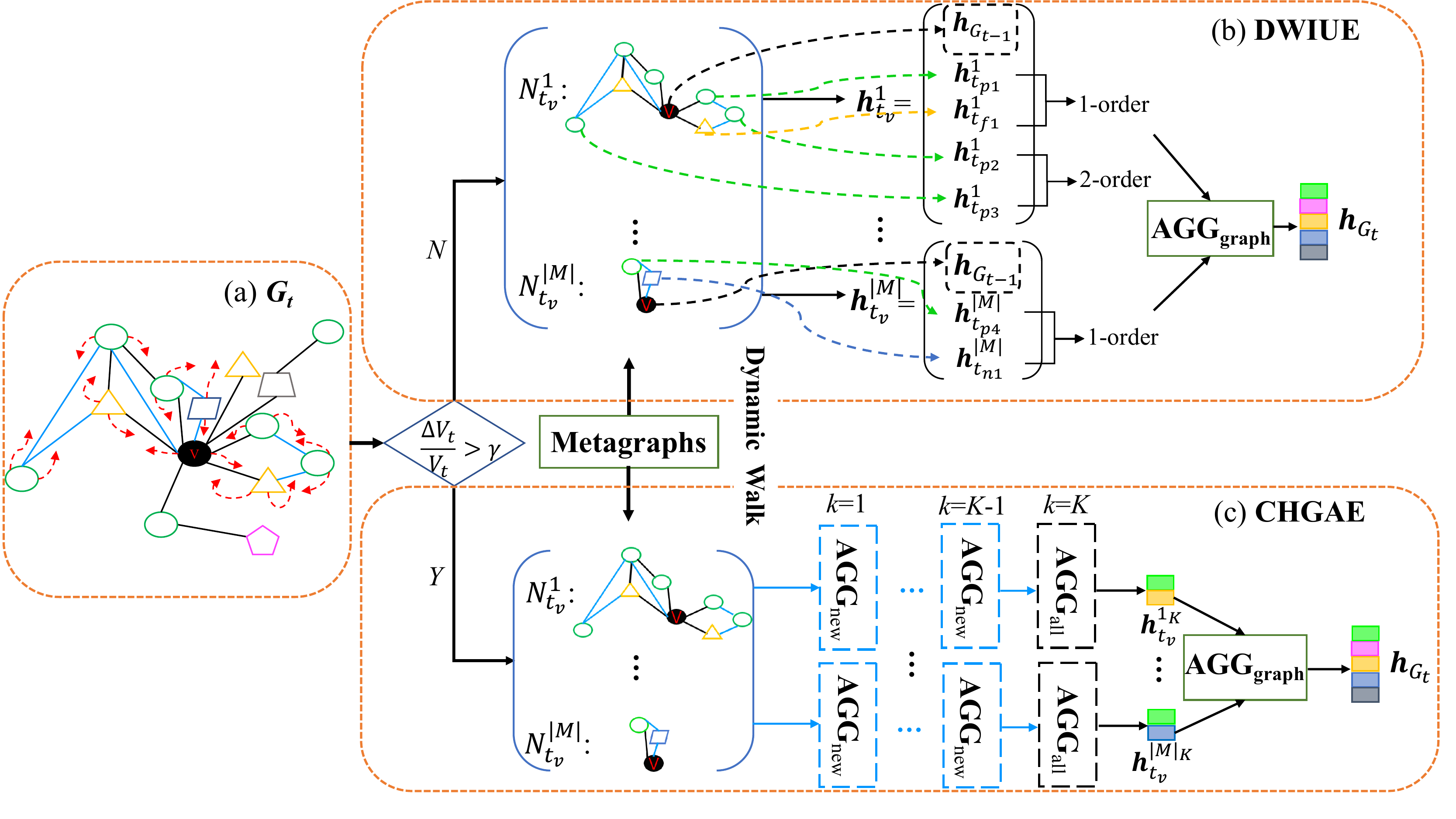}  
    \caption{Dynamic graph learning: (a) Heterogeneous graph {\em G$_{t}$} at sliding window {\em W$_{t}$}, where the dotted lines with the red arrows indicate the dynamic nodes that need to perform random walks; (b) Dynamic Walk Incremental Updating Encoder (DWIUE); (c) Combined Hierarchical Graph Attention Encoder (CHGAE), where {\em AGG$_{new}$}, {\em AGG$_{all}$}, and {\em AGG$_{graph}$} denote dynamic node aggregator, node aggregator, and meta-graph aggregator, respectively.}  
    \label{fig:method}
\end{figure*}

\subsubsection{Dynamic Walk Incremental Updating Encoder}
As mentioned above, DWIUE focuses on dynamically updating graph embedding when {\em $\Delta V_{t}$}/{\em V$_{t}$}$\leq\gamma$, which involves three major steps: (A) Construct dynamic walk neighbor set; (B) Incrementally fuse representation matrix; and (C) Aggregate graph embedding.

{\bf (A) Construct dynamic walk neighbor set {\em N$_{t_{v}}^{i}$}:} Given the heterogeneous graph {\em G$_{t}$} and the meta-graphs {\em M=$\{$M$_{1}$, M$_{2}$,\dots, M$_{|M|}\}$}, the path-relevant neighbor set {\em N$_{t_{v}}^{i}$} of the target process $v$ can be obtained with the guidance of each {\em M$_{i}$}.

{\bf (B) Incrementally fuse representation matrix {\bf \em h$_{t_{v}}^{i}$}:}
As shown in Figure~\ref{fig:method}(b), a representation matrix of the target process $v$ can be established by fusing {\bf \em h$_{G_{t-1}}$} at %the previous window
{\em W$_{t-1}$} and its new neighbors' feature vectors at {\em W$_{t}$}, which leverages both the embedding of itself and the information of its new neighbors in {\em N$_{t_{v}}^{i}$}. 
Innovatively, DWIUE selects the intersection between {\em N$_{t_{v}}^{i}$} and {\em $\Delta V_{t}$} as the meaningful dynamic neighbors, and the representation matrix {\bf \em h}$_{t_{v}}^{i} \in \mathcal{R}^{d\times d}$ is expressed as:
\begin{equation}
      {\bf h}_{t_{v}}^{i}=[{\bf h}_{G_{t-1}}, {\bf h}_{t_{u1}}^{i},\dots, {\bf h}_{t_{un}}^{i}]^{T},\label{eq:no2}
\end{equation}
where {\em u1~$\neq$~un}~$\in$~({\em N$_{t_{v}}^{i}\wedge\Delta V_{t}$}), and they are sorted by the node types~(i.e., process, file, memory, registry, system, mutex, attribute, and network.) and the distance to node {\em v}.

{\bf (C) Aggregate graph embedding {\bf \em h$_{G_{t}}$}:}
Actually, different meta-graphs show distinctive contributions in characterizing malware variants. Inspired by~\cite{DBLP:conf/ijcai/WangCYLNTGLCY19}, we implement a self-attention mechanism to learn the weight of each meta-graph for detecting malware variants. Concretely, the attention weight {\em $\theta_{t}^{i}$} of {\em $M_{i}$} is formalized as below.
\begin{equation}
\resizebox{.91\linewidth}{!}{$
    \displaystyle
      \theta_{t}^{i}= \frac{exp(LeakyReLU({\bf W}^{l}[{\bf h}_{t_{v}}^{i},{\bf h}_{t_{v}}^{j}]+b^{l}))}{\sum_{m\in |M|}exp(LeakyReLU({\bf W}^{l}[{\bf h}_{t_{v}}^{i},{\bf h}_{t_{v}}^{m}]+b^{l}))},\label{eq:no3}
      $}
\end{equation}
where {\em i $\neq$ j} $\in\{$1,\dots, $|\emph{M}|\}$, {\bf \em h$_{t_{v}}^{i}$} and {\bf \em h$_{t_{v}}^{j}$} denote the node representations of the target process node {\em v} under meta-graph {\em M$_{i}$} and {\em M$_{j}$}, respectively. {\bf \em W$^{l}$} and {\em b$^{l}$} are the trainable weight and bias parameters.
Naturally, the graph embedding of {\em $G_t$} with different meta-graphs can be aggregated as follows.
\begin{equation}
      {\bf h}_{G_{t}}=\sum_{i=1}^{|M|}\theta_{t}^{i}\times {\bf h}_{t_{v}}^{i}.\label{eq:no4}
\end{equation}

\subsubsection{Combined Hierarchical Graph Attention Encoder}
As shown in Figure~\ref{fig:method}(c), CHGAE is presented to handle the graph embedding of $G_t$ when threshold {\em $\Delta V_{t}$}/{\em V$_{t}$}$>\gamma$. Particularly, CHGAE also consists of three key steps: (A) Construct dynamic walk neighbor set; (B) Aggregate node embedding; (C) Aggregate graph embedding. Notably, steps (A) and (C) of CHGAE are similar to DWIUE. Here, we detail step (B), which can discriminatively aggregate newly joined and known nodes.

{\bf (B) Aggregate node embedding {\bf \em h$_{t_{v}}^{i_{K}}$}:}
CHGAE calculates the significance of neighbor nodes for representing the target process node {\em v}, which specifically selects more important neighbor nodes to represent itself rather than uniformly aggregating its neighbor nodes in {\em N$_{t_{v}}^{i}$}. Formally, the weight of the neighbor node can be defined as:
\begin{equation}
      \alpha_{t_{u}}^{i}= \frac{exp(\frac{|N_{t_{v}}^{i}.u[type]|}{|N_{t_{v}}^{i}|\times n})}{\sum_{u\in N_{t_{v}}^{i}}exp(\frac{|N_{t_{v}}^{i}.u[type]|}{|N_{t_{v}}^{i}|\times n})},\label{eq:no5}
\end{equation}
where {\em N$_{t_{v}}^{i}.u[type]$} denotes the node set whose type is the same as instance {\em u} in {\em N$_{t_{v}}^{i}$}, and {\em n} depicts {\em u} as the {\em n}-order neighbor of the target process node {\em v}.

{\em G$_{t}$} evolves from the {\em $DHGS=\{G_1, G_2,\dots, G_{t-1}\}$}, which contains the previously known nodes before {\em $W_t$} and newly added nodes at {\em $W_t$}. Notably, the known nodes in {\em G$_{t}$} have been aggregated in {\em G$_{t-1}$}; thus, CHGAE only learns newly added nodes in {\em N$_{t_{v}}^{i}$} at {\em $W_t$}. As shown in Figure~\ref{fig:method}(c), the former ({\em K}-1) layers of the proposed aggregator only tackle the new neighbors in {\em N$_{t_{v}}^{i}$}, and the last layer aggregates all neighbors in {\em N$_{t_{v}}^{i}$}, which greatly reduces the time cost of node embeddings. The former ({\em K}-1) layers can be defined as:

\begin{align}
      {\bf h}_{t_{v}}^{i_{k}} &=AGGREGATE_{k}({\bf h}_{t_{v}}^{i_{k-1}}, \{{\bf h}_{t_{u}}^{i_{k-1}}\}_{u\in (N_{t_{v}}^{i}\wedge\Delta V_{t})})\nonumber\\
     &=\sigma((1+\epsilon_{k}) {\bf h}_{t_{v}}^{i_{k-1}}+\sum_{u\in (N_{t_{v}}^{i}\wedge\Delta V_{t})} \alpha_{t_{u}}^{i}{\bf W}_{k}{\bf h}_{t_{u}}^{i_{k-1}}),\label{eq:no6}
\end{align}
where {\em k} $\in \{$1, 2,\dots, {\em K}-1$\}$ denotes the index of the layer, {\bf \em h$_{t_{v}}^{i_{k}}$} and {\bf \em h$_{t_{u}}^{i_{k}}$} are the feature vectors of target process node {\em v} and neighbor node {\em u} at the {\em k$_{th}$} layer, respectively, and they can initialize by their state vectors. {\em $\epsilon_{k}$} is a trainable trade-off parameter. Moreover, the {\em K$_{th}$} layer of the proposed aggregator is similar to the previous ones, which aggregates all neighbors in {\em N$_{t_{v}}^{i}$} to represent the final node embedding. 

\begin{table}[t]
    \small
	\centering
	\begin{tabular}{lrlrlr}
	%\begin{tabular*}{\hsize}{@{}@{\extracolsep{\fill}}cc|cc@{}}
	%\begin{tabular*}{\hsize}{@{}@{\extracolsep{\fill}}cc@{}}
		\toprule
		Type & Samples & Type & Samples & Type & Samples\\
	\midrule
		Trojan & 4,536 & Virus & 1,606& Worm & 842\\
		Backdoor & 660 & Adware & 394 & Exploit & 338\\
		Dropper & 118 & Benign & 3,042 \\ 
	\bottomrule
		%\end{tabular*}
	%\end{tabular*}
		\end{tabular}
		\caption{Sample distribution in ACT-KingKong dataset.}
		\label{sample}
\end{table}
\begin{table}[t]
    %\small
	\centering
	\begin{tabular}{lrlr}
		\toprule  
		Entity  &  Number  &  Entity  &  Number \\
		\midrule
		Process  &  2,869,361  &  File  &  1,284,625 \\
		Memory  &  641,945  &  Registry  &  436,773 \\
		System  &  284,614  &  Mutex  &  44,498 \\ 
		Attribute  &  31,608  &  Network  &  26,976 \\
	\bottomrule
	\end{tabular}
		\caption{Statistics of ACT-KingKong dataset.}
		\label{ststistics}
\end{table}
\subsection{Real-time Detection}
In this section, we leverage the Pearson correlation coefficient~\cite{feng2019expert} between the target variant and all samples to detect the target variant in real time. Intuitively, the real intentions~(e.g., stealing sensitive information, etc.) of the new variant is similar to that of the original malware. Formally, the Pearson correlation coefficient is defined as:
\begin{align}
     \rho({\bf h}_{G_{t}},{\bf h}_{G_{f}}) 
     &=\frac{E\lbrack ({\bf h}_{G_{t}}-\mu_{{\bf h}_{G_{t}}})({\bf h}_{G_{f}}-\mu_{{\bf h}_{G_{f}}})\rbrack}{\sigma_{{\bf h}_{G_{t}}}\sigma_{{\bf h}_{G_{f}}}},\label{eq:no7}
\end{align}
where {\em $\mu_{{\bf h}_{G_{t}}}$} and {\em $\mu_{{\bf h}_{G_{f}}}$} are expectations of {\bf \em h$_{G_{t}}$} and {\bf \em h$_{G_{f}}$}, respectively. {\em $\sigma_{{\bf h}_{G_{t}}}$} and {\em $\sigma_{{\bf h}_{G_{t}}}$} are standard deviations of {\bf \em h$_{G_{t}}$} and {\bf \em h$_{G_{f}}$}, respectively. 

MG-DVD outputs the type of the sample with the highest Pearson correlation coefficient exceeded the threshold {\em $\tau$} as the detection result. Particularly, if {\em $\rho$({\bf \em h$_{G_{t}}$}, {\bf \em h$_{G_{f}}$)$\geq\tau$}}, MG-DVD continues to slide forward until the two detection results are consistent. Inversely, if the Pearson correlation coefficients with all samples are less than {\em $\tau$}, it is essential to slide forward and learn a richer embedding to continue detecting.

Finally, the objective function of dynamic malware variants detection is formulated as below.
\begin{equation}
     l=\sum_{p=1}^{|P|}(\rho({\bf h}_{G_{p1}},{\bf h}_{G_{p2}})-y_{p})^{2},\label{eq:no8}
\end{equation}
where {\em y$_{p}$}=1 if {\bf \em h$_{G_{p1}}$} and {\bf \em h$_{G_{p2}}$} are graph embeddings from the same malware; otherwise, label {\em y$_{p}$}=-1. Here, we optimize {\em l} with Adam optimizer~\cite{kingma2014adam}.

\section{Experiments}
\subsection{Dataset}
We collect a large number of real-world executable (PE) files from VirusTotal,\footnote{\url{https://www.virustotal.com.}} which contains the latest Windows malware and variants from Mar 2019 to Oct 2019.
Afterwards, the execution events streams dataset of all PE files is captured using the KingKong system, namely ACT-KingKong.\footnote{\url{https://github.com/yidun1027/ACT-KingKong.}} More specifically, the sample distribution and basic statistics for our ACT-KingKong dataset are presented in Table~\ref{sample} and Table~\ref{ststistics}, respectively. For all samples, we randomly divide the training set, validation set, and test set into 6:2:2.

\begin{table}[t]
    \small
	\centering
	\begin{tabular}{lrrrrrr}
		\toprule  
		Method & Recall & Precision & ACC &
		F-1 & AUC \\ 
		\midrule 
	SVM+RBF & 0.767 & 0.878 & 0.764 & 0.819 & 0.701 \\
	RNN+LR & 0.847 & 0.911 & 0.851 & 0.878 & 0.868 \\
	MalConv & 0.833 & 0.908 & 0.843 & 0.869 & 0.839  \\
	CNN+BPNN & 0.873 & 0.919 & 0.882 & 0.895 & 0.884 \\
	MatchGNet & 0.913 & 0.937 & 0.917 & 0.925 & 0.916\\
    {\bf MG-DVD} & {\bf 0.965} & {\bf 0.981} & {\bf 0.976} & {\bf 0.973} & {\bf 0.952}\\
		\bottomrule 
	\end{tabular}
	\caption{Performance on malware variants detection.}
	\label{metric}
\end{table}
\begin{figure}[t]  
    \centering 
    \includegraphics[width=5.7cm]{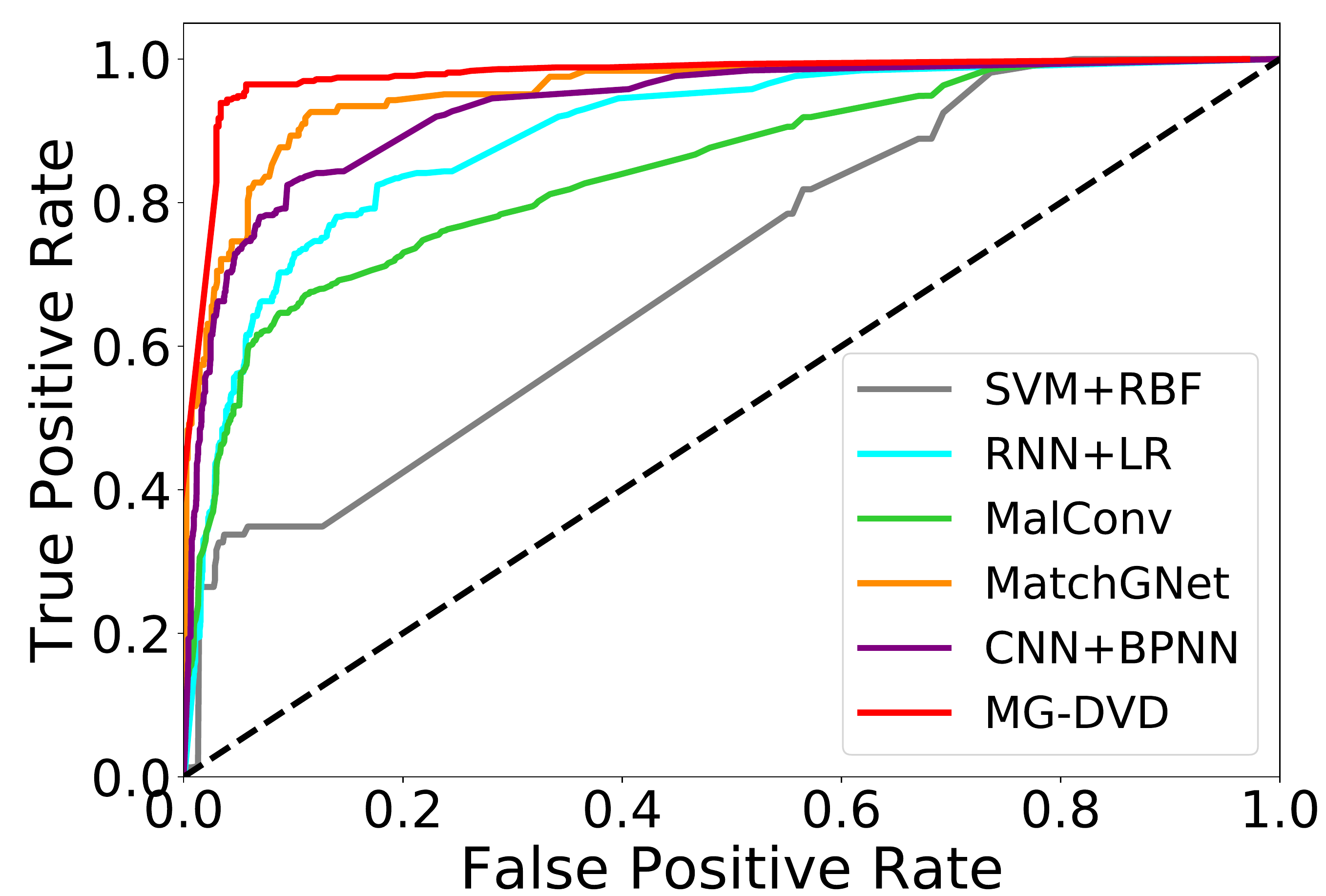} 
    \caption{ROC curves on malware variants detection.}  
    \label{roc}
\end{figure}
\subsection{Result Analysis}
In this section, we validate the effectiveness, efficiency, and interpretability of MG-DVD in detecting malware variants by comparing it with five strong baseline methods, where SVM+RBF~\cite{gaviria2017using}, MalConv~\cite{raff2018malware} and CNN+BPNN~\cite{zhang2019feature} belong to signature-based detection methods, RNN+LR~\cite{pascanu2015malware} and MatchGNet~\cite{DBLP:conf/ijcai/WangCYLNTGLCY19} belong to behavior-based detection methods. We implement or utilize the source codes shared by the authors and adopt the same parameters in their works.
\begin{table*}[t]
    \small
	\centering
	\setlength{\tabcolsep}{1.7pt}
	\begin{tabular}{lrrrrrrrr}
		\toprule  
		%Method & Total time(s) & Detection time(s) \\ 
		Method &SVM+RBF &RNN+LR &MalConv &CNN+BPNN &MatchGNet &MG-DVD$_{CHGAE}$ &MG-DVD$_{static walk}$ &{\bf MG-DVD}\\
		\midrule 
		 Total time(s) &108.13 & 156.44 &79.78 &202.53 &414.35 & 1083.07 &1035.81 &{\bf 345.50}\\
		Detection time(s) & 5.60 & 3.37&  3.21& 7.12&  13.79& 25.21& 19.33&  {\bf 8.84}\\ 
		\bottomrule  
	\end{tabular}
		\caption{Runtime comparison.}
		\label{runtime}
\end{table*}
\subsubsection{Effectiveness Evaluation}
As shown in Table~\ref{metric} and Figure~\ref{roc}, our proposed MG-DVD outperforms all baseline methods in terms of all metrics. The improvement of MG-DVD can be attributed to the following traits. First, comparing with feature-based methods~(e.g., SVM+RBF, MalConv, RNN+LR, and CNN+BPNN), MG-DVD models the fine-grained execution events streams of malware variants into dynamic heterogeneous graphs sequences to effectively learn the evolutionary patterns between malware and their variants. Notably, our MG-DVD achieves a 9–21\% improvement in terms of ACC against feature-based methods. Second, compared with MatchGNet, MG-DVD implements a metagraph-guided dynamic graph learning method, which can learn more fine-grained representations of malware variants to accurately distinguish the new malware variants. Indeed, MG-DVD generates fewer false positives~(i.e., 21) than MatchGNet~(i.e., 46), indicating that MG-DVD can relieve alarm fatigue.
\subsubsection{Efficiency Evaluation}
Here we evaluate the efficiency of MG-DVD from two aspects. Firstly, we compare it with its two variants, namely MG-DVD$_{static walk}$ and MG-DVD$_{CHGAE}$. As shown in Table~\ref{runtime}, MG-DVD is 2.18$\times$ and 3$\times$ faster than MG-DVD$_{static walk}$ in terms of detection time and total time. Moreover, MG-DVD is significantly faster than MG-DVD$_{CHGAE}$ since the encoder DWIUE in MG-DVD can skillfully reduce the time for node aggregation compared to the CHGAE. Secondly, we compare MG-DVD with the existing feature-based and metapath-based methods. As we presented in Table~\ref{runtime}, feature-based detection methods, such as SVM+RBF, RNN+LR, MalConv, and CNN+BPNN, are faster than MG-DVD due to the fact that they do not consider the complex graph structure from various types of entity and relation. On the contrary, the static MatchGNet is slower than MG-DVD, which means that the metagraph-guided dynamic embedding encoders in our MG-DVD are beneficial to real-time malware variant detection.
\begin{figure}[t]  
    \centering 
    \includegraphics[width=6cm]{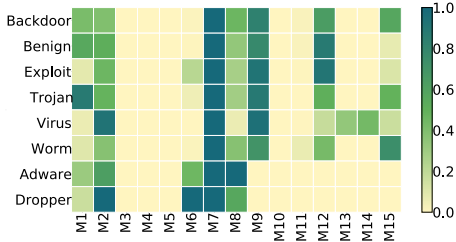} 
    \caption{Interpretability for MG-DVD.}  
    \label{interpretability}
\end{figure}

\subsubsection{Interpretability Evaluation}
Figure~\ref{interpretability} shows the frequency matrix of the top-{\em k} meta-graphs with the largest weights for different types of malware, which improves interpretability for malware variants detection, from which we can learn that: (i) several meta-graphs show crucial significance on detecting all types of malware variants, such as {\em M$_{1}$}, {\em M$_{2}$}, {\em M$_{7}$}, and {\em M$_{8}$}; (ii) several meta-graphs only focus on specific types, such as {\em M$_{13}$} and {\em M$_{14}$} are only involved in Virus, which depicts that the Virus is usually active in connecting to the network and infecting files.

\subsection{Parameter Sensitivity}
In this section, we conduct a large volume of experiments to analyze the sensitivity of different parameters in MG-DVD, including the number of layers~(i.e., {\em K}), dimension of representation matrix~(i.e., {\em d}), number of hidden neurons~(i.e., {\em N}), embedding size~(i.e., {\em E}), window size~(i.e., {\em W}), and sliding step~(i.e., {\em p}).

\begin{figure}[t]  
    \centering  
    \subfigure[\emph{K} and \emph{d} vs. ACC ]{
        \begin{minipage}[t]{0.5\linewidth}
            \centering  
            \includegraphics[width=0.8\linewidth]{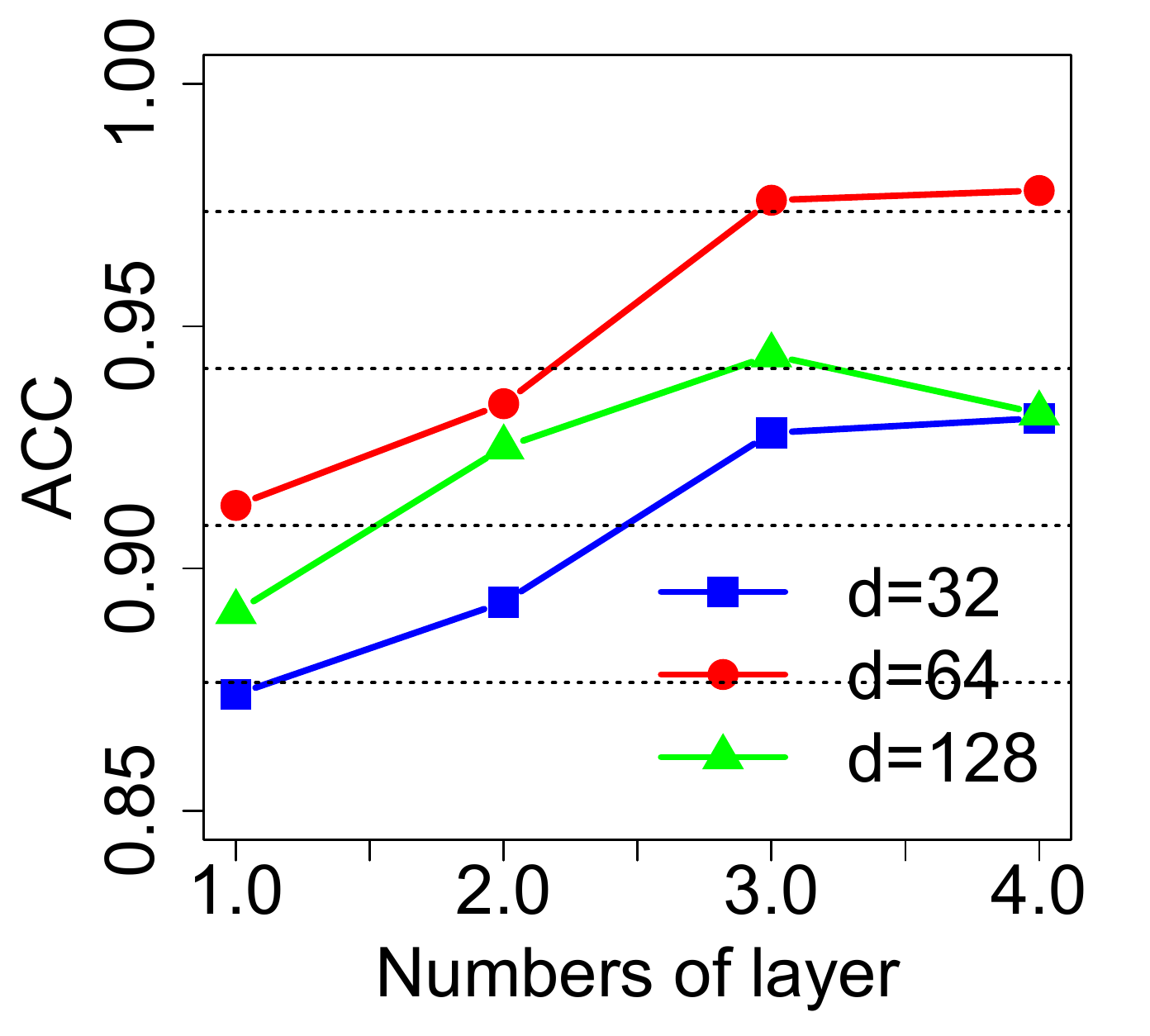}  \end{minipage}
}\subfigure[\emph{N} and \emph{E} vs. ACC]{
        \begin{minipage}[t]{0.5\linewidth}
            \centering     
            \includegraphics[width=0.82\linewidth]{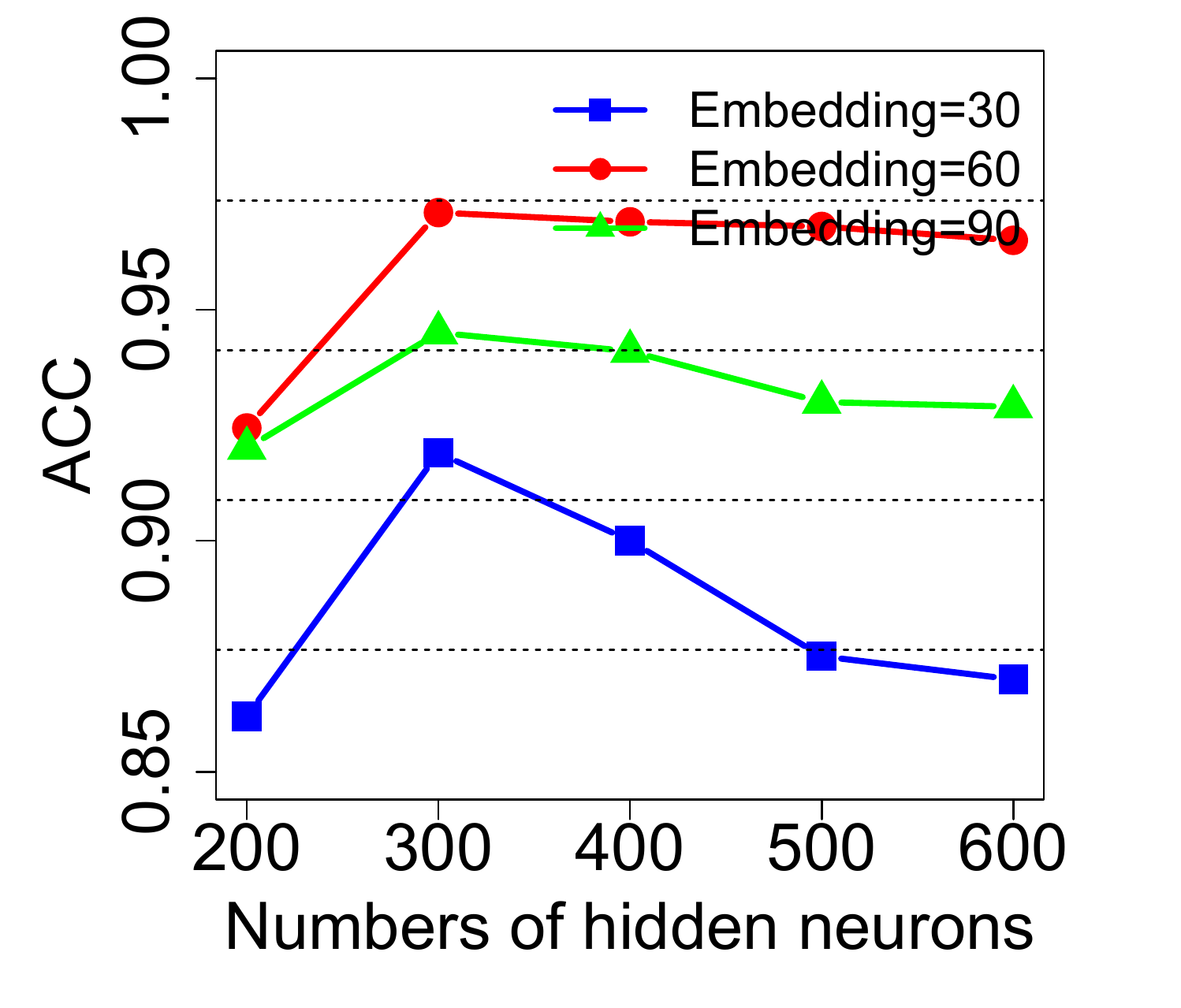}  \end{minipage}
    }
    
\subfigure[{\em W} and {\em p} vs. detection time]{
        \begin{minipage}[t]{0.5\linewidth}
            \centering  
            \includegraphics[width=0.85\linewidth]{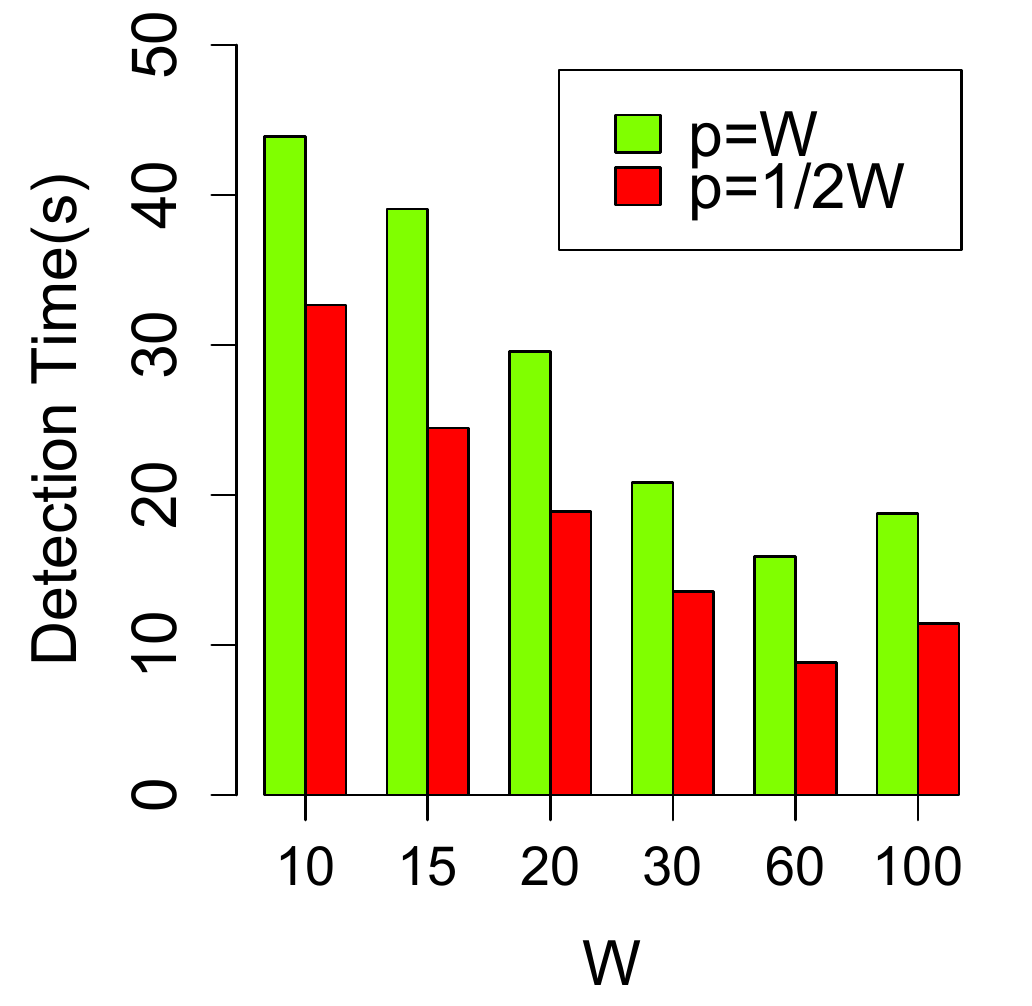}   
        \end{minipage}
}\subfigure[{\em W} and {\em p} vs. ACC]{
        \begin{minipage}[t]{0.5\linewidth}
            \centering     
            \includegraphics[width=0.78\linewidth]{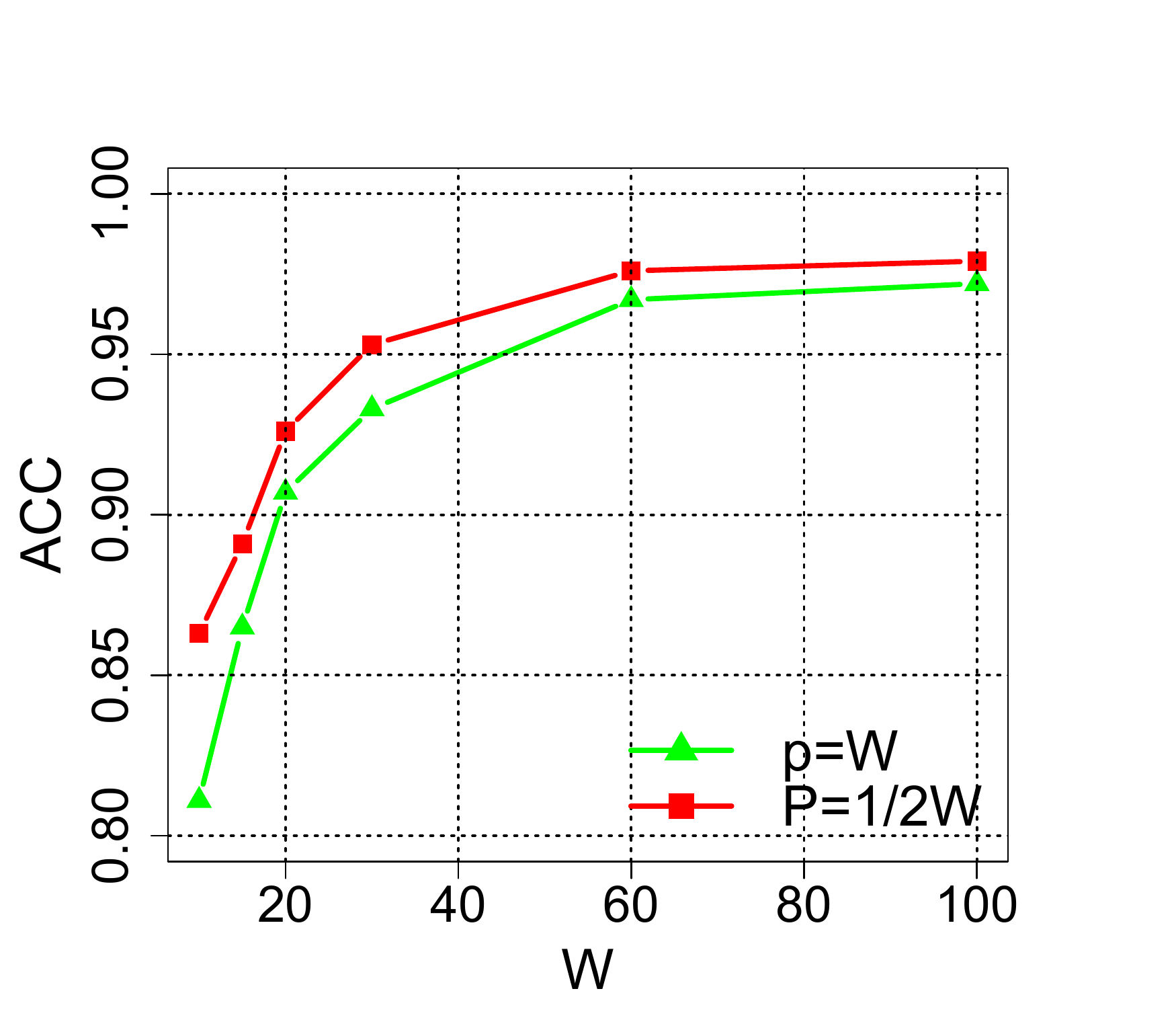}   
        \end{minipage}
    }  
    \caption{Parameter sensitivity.} 
    \label{fig1} 
\end{figure}

As shown in Figure~\ref{fig1}(a), ACC stably increases with {\em K} increases from 1 to 3, when {\em K}= 4, the framework is more complex, yet has little improvement on the ACC. Moreover, the performance of our MG-DVD is outstanding in terms of accuracy and stability when the dimension of the representation matrix is set to 64. Similarly, as shown in Figure~\ref{fig1}(b), our MG-DVD is the most stable and effective when the number of hidden neurons is set to 300 and embedding size = 60. 
As demonstrated in Figure~\ref{fig1}(c), the detection time when {\em p}=1/2{\em W} is shorter than that of {\em p}={\em W}, which can be attributed to adjacent sliding windows having more overlapping behaviors when {\em p}=1/2{\em W}, reducing the time cost of node retraining. Moreover, the larger {\em W}, the less detection time; however, when {\em W} is greater than the 60s, the detection time becomes longer because the number of sliding times is fixed when {\em W} reaches a certain value~(i.e., 60s), but the more neighbors need to walk and aggregate in larger windows, causing more time cost. In addition, in Figure~\ref{fig1}(d), we can observe that the improvement of ACC becomes slow after {\em W}=60s. Consequently, we set {\em W}=60s and {\em p}=1/2{\em W} in MG-DVD to hold the trade-off between detection accuracy and efficiency.

\subsection{Ablation Study}
\begin{table}[t]
    \small
	\centering
	\setlength{\tabcolsep}{0.2pt}
	\begin{tabular}{llllr}
		\toprule  
		Method &Walk &Embedding &Meta-structure &ACC\\
		\midrule 
		 MatchGNet &static & HAGNE &meta-path & 0.917\\
		MG-DVD$_{static walk}$ &static & DWIUE+CHGAE &meta-graph & 0.958\\
		MG-DVD$_{CHGAE}$ &dynamic & CHGAE &meta-graph &0.924 \\
		MG-DVD$_{DWIUE}$ &dynamic & DWIUE &meta-graph & 0.853\\
		MG-DVD &dynamic & DWIUE+CHGAE &meta-graph & 0.976\\
		\bottomrule  
	\end{tabular}
		\caption{A summary of the ablation study.}
		\label{ablation}
\end{table}
We perform a detailed ablation study comparing different variants of our MG-DVD, investigating the effects of different modules in the model. Table~\ref{ablation} summarizes our experiments. We first evaluate the two metagraph-guided dynamic embedding encoders by comparing MG-DVD with MatchGNet~\cite{DBLP:conf/ijcai/WangCYLNTGLCY19}, which indicates the importance of the meta-graph and two dynamic embedding encoders~(i.e., DWIUE and CHGAE) to improve detection performance. We also study the effect of the dynamic walk strategy by comparing MG-DVD with MG-DVD$_{static walk}$ that only adopted static random walk strategy. Table~\ref{ablation} shows that the dynamic walk strategy in MG-DVD outperforms the existing static random walk strategy since the dynamic walk strategy only traverses the newly added nodes rather than all nodes in the target graph. Additionally, we examine the discriminative effects of each metagraph-guided dynamic embedding encoder by experimenting with MG-DVD$_{CHGAE}$ and MG-DVD$_{DWIUE}$, respectively. As can be seen from Table~\ref{ablation}, the detection accuracy of MG-DVD$_{CHGAE}$ is a significant boost than that of MG-DVD$_{DWIUE}$, which can be attributed to employing aggregators to aggregate high-order neighborhoods in CHGAE can more effectively capture semantic information than directly expanding the state vector of the target node in DWIUE.  
\section{Related Work}
\subsection{Signature-based Malware Detection}
There have been numerous studies concerning signature-based malware detection~\cite{gandotra2014malware,kang2016n,gaviria2017using,raff2018malware,zhang2019feature,singh2020survey}. Concretely, \cite{kang2016n,gaviria2017using,raff2018malware} extracted the opcode or bytecode to detect malware, yet they rely on manually extracting features from known malware samples, which hardly identify new types of malware. Zhang et al.~\cite{zhang2019feature} presented a hybrid model consisting of CNN and BPNN to extract more advanced malware features from raw opcodes and APIs. However, they are vulnerable to code obfuscation techniques, inevitably leading to a drop in malware variants detection performance.

\subsection{Behavior-based Malware Detection}
Recently, extensive studies were implemented on detecting malware based on execution behaviors~\cite{pascanu2015malware,sun2016monet,bartos2016optimized,zhang2018sensitive,DBLP:conf/ijcai/WangCYLNTGLCY19,zhang2020dynamic}. Pascanu et al.~\cite{pascanu2015malware} and Zhang et al.~\cite{zhang2018sensitive} verified that the API sequences of malware are significantly different from that of the legitimate programs, which provides a new avenue for behavior-based malware detection. Nevertheless, these methods ignored the interactive relationships among various malware objects and cannot learn the evolutionary patterns between malware and their variants.

To address the limitations, Wang et al.~\cite{DBLP:conf/ijcai/WangCYLNTGLCY19} proposed MatchGNet,
and they characterized execution events of malware into a static heterogeneous graph and extracted metapath-based features to detect malware, which can effectively identify the unknown malware from a large number of benign samples.
However, MatchGNet exposes two serious deficiencies. First, it focused on learning metapath-based malware features, which is unable to capture high-order and fine-grained malware variant patterns, resulting in a drop in malware variants detection performance and high false positive. Secondly,
MatchGNet learned a long-term representation from the static heterogeneous graph, which cannot meet the
dynamic detection requirements.

\section{Conclusion}
In this paper, we proposed MG-DVD, a dynamic malware variants detection framework. MG-DVD first utilized dynamic heterogeneous graphs sequence~(DHGS) to model the execution events streams of malware variants and introduced meta-graphs to characterize the interactive relationships between malware objects. To realize real-time malware variants detection, MG-DVD then proposed two dynamic walk-based encoders, which leverages overlapping information between adjacent sliding windows to dynamically learn node embeddings. Finally, MG-DVD trained a Pearson-based model to automatically detect malware variants. Extensive experiments show that MG-DVD is effective and efficient in dynamically detecting new malware variants. In the future, we would expand the MG-DVD to a more high-level dynamic detection method by considering the temporal correlation between various behaviors.

\section*{Acknowledgements}
This work was supported by the 2020 Industrial Internet Innovation and Development Project-the Key Project of Intelligent Connected Vehicle Safety Inspection Platform~(Tender No. TC200H01S), the Opening Project of Shanghai Trusted Industrial Control Platform (TICPSH202003020-ZC), and the Beijing Advanced Innovation Center for Big Data and Brain Computing.

%% The file named.bst is a bibliography style file for BibTeX 0.99c
\bibliographystyle{named}
%\bibliography{ijcai21}

%\bibliographystyle{IEEEtran}
%\bibliography{ijcai21}

\end{document}